\documentclass{svproc}
\usepackage{url}
\usepackage{graphicx}
\usepackage[english, brazil]{babel} % for European Portuguese use portuguese
\usepackage[fixlanguage]{babelbib}

\begin{document}
\mainmatter              % start of a contribution
\title{Software multiplataforma para a segmentação de vasos sanguíneos em imagens da retina}
\titlerunning{Software de segmentação de vasos sanguíneos retinais}  % abbreviated title (for running head)
\author{João Henrique Pereira Machado\inst{1}, Gilson Adamczuk Oliveira\inst{1} \and Érick Oliveira Rodrigues\inst{1}
}
\authorrunning{João Henrique Pereira Machado et al.} % abbreviated author list (for running head)
\tocauthor{João Henrique Pereira Machado and Érick Oliveira Rodrigues}
\institute{Universidade Tecnológica Federal do Paraná (UTFPR), Pato Branco, Brasil,\\
\email{joaohenriquemachado@alunos.utfpr.edu.br},\email{erickrodrigues@utfpr.edu.br}\\
}

\maketitle              % typeset the title of the contribution

\begin{abstract}
Neste trabalho, utilizamos a segmentação de imagens para identificar visualmente vasos sanguíneos em imagens de exames de retina. Este processo normalmente é realizado manualmente. Contudo, podemos utilizar métodos heurísticos e aprendizado de máquina para realizar o processo de forma automatizada, ou ao menos acelerá-lo. Nesse contexto, propomos um software multi-plataforma, open-source e responsivo, que permite que o usuário segmente manualmente uma imagem de retina, com o intuito de utilizar a imagem segmentada pelo usuário para retreinar algoritmos de aprendizado de máquina para melhorar futuros resultados de segmentações automáticas. Além disso, o software também é capaz de implementar e disparar alguns filtros de imagem que já estão dispostos na literatura, para melhorar a visualização dos vasos. Propomos a primeira solução do tipo na literatura. Este é o primeiro software integrado que possui as características levantadas: open-source, responsivo e multi-plataforma, sendo uma solução completa desde realizar a segmentação dos vasos manualmente, até disparar de forma automática algoritmos de classificação para aprimoramento dos modelos preditivos.
\end{abstract}

\section{Introdução}
Segmentar uma imagem digital pode ser interpretado como encontrar suas regiões homogêneas e suas bordas, ou limites. As regiões homogêneas devem corresponder a partes significativas de objetos do mundo real, e as bordas aos seus contornos aparentes. \cite{morel2012variational}.

Na oftalmologia, a segmentação de imagens é importante, por exemplo, na identificação de doenças que que estão ligadas à expansões ou reduções na estrutura da vasculatura e no processo e análise de imagens retinais. A segmentação pode contribuir no registro de imagens de retina \cite{Rodrigues2016} (por exemplo, em fluoroscopia de vídeo)\cite{Rodrigues2018}, na localização do disco óptico e da fóvea e auxiliar na análise de fluxo sanguíneo \cite{VargasCanas2012}.

Por exemplo, aneurismas são caracterizados por dilatações anormais de uma artéria. Essa complicação pode ser encontrada como consequência do diabetes chamada retinopatia diabética, causando danos aos vasos sanguíneos da retina. 

\begin{figure}[h]
\centering
\includegraphics[width=\columnwidth]{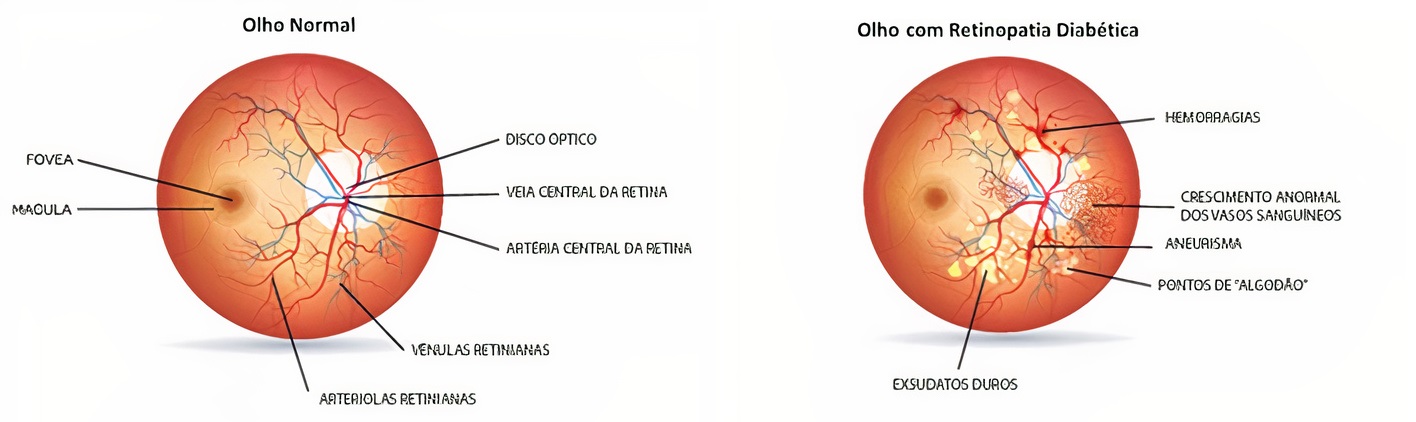}
\caption{Diferenças entre um olho saudável e o olho de uma pessoa com retinopatia diabética \cite{retinopatia}.}
\label{fig:aneurisma}
\end{figure}

Ao destacar a vasculatura da retina, é possível detectar este tipo de anormalidade observando as regiões com expansões espontâneas do vaso sanguíneo. Esse efeito é demonstrado na Figura \ref{fig:aneurisma}.

Outro tipo de anormalidade são estenoses, caracterizadas pelo estreitamento de um vaso sanguíneo. Reduções repentinas e inesperadas na área segmentada podem indicar restrições no fluxo sanguíneo \cite{Rodrigues2020}.

\begin{figure}[h]
\centering
\includegraphics[width=9.5cm]{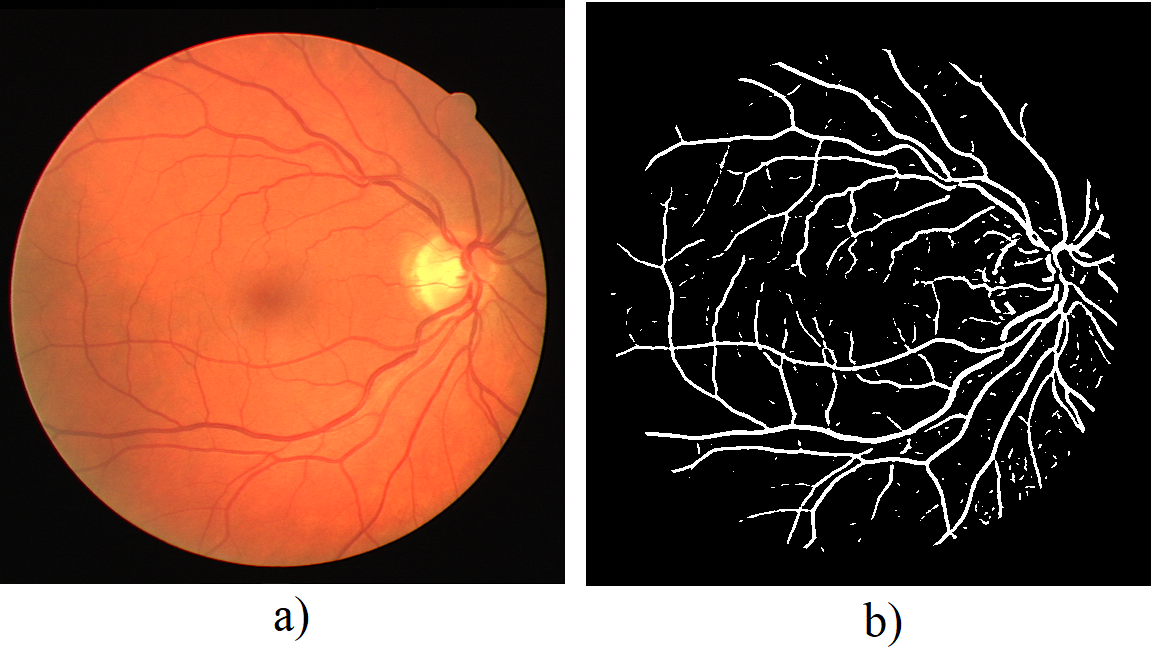}
\caption{Em (a) uma imagem de entrada e em (b) o resultado da aplicação de um filtro de segmentação.}
\label{fig:frangi_example}
\end{figure}

Na Figura \ref{fig:frangi_example} temos um exemplo da aplicação da segmentação em uma imagem de retina, destacando a vasculatura do olho.

A proposta deste software tem como objetivo auxiliar a aplicação de ferramentas de segmentação no ambiente médico e o desenvolvimento e aprimoramento de métodos computacionais de segmentação.

O software foi desenvolvido utilizando a linguagem Java, para facilitar o desenvolvimento para diferentes plataformas. A biblioteca LibGDX foi utilizada para criar a interface gráfica, garantindo um ambiente acessível, responsivo e idêntico, independente da plataforma utilizada. 

O framework Weka \cite{Weka} foi utilizado para construção e treinamento dos modelos de classificação, além da aplicação dos mesmos, que podem ser utilizados para segmentar novas imagens.

\section{Desenvolvimento do Software}
O software proposto possui três ferramentas principais:

\begin{itemize}
    \item Edição de segmentações: um editor para a criação e edição de segmentações, similar a ferramentas como Photoshop e Paint; com intuito de auxiliar o treinamento com as novas correções para gerar melhores segmentações automáticas;
    \item Algoritmos de segmentação: filtros de segmentação que criam uma segmentação da imagem utilizando técnicas de processamento de imagem;
    \item Classificadores e modelos preditivos: classificadores que criam modelos de predição e podem ser utilizados para segmentar a imagem, avaliando padrões nas segmentações.
\end{itemize}

O software possui a versatilidade para trabalhar com diferentes modalidades de imagens e situações, com a finalidade de aproveitar as melhores características de abordagem.

\subsection{Editor de segmentações}
Primeiramente, o software possui um editor de imagens para criar e editar segmentações. É similar a outros editores de imagem como Photoshop e Paint. Entretanto, a imagem é separada em camadas, com a imagem original na camada mais interna, enquanto a segmentação é apresentada em uma camada mais externa. Assim se torna bastante prático extrair e trabalhar com a segmentação da imagem.

\begin{figure}[h]
\centering
\includegraphics[width=\columnwidth]{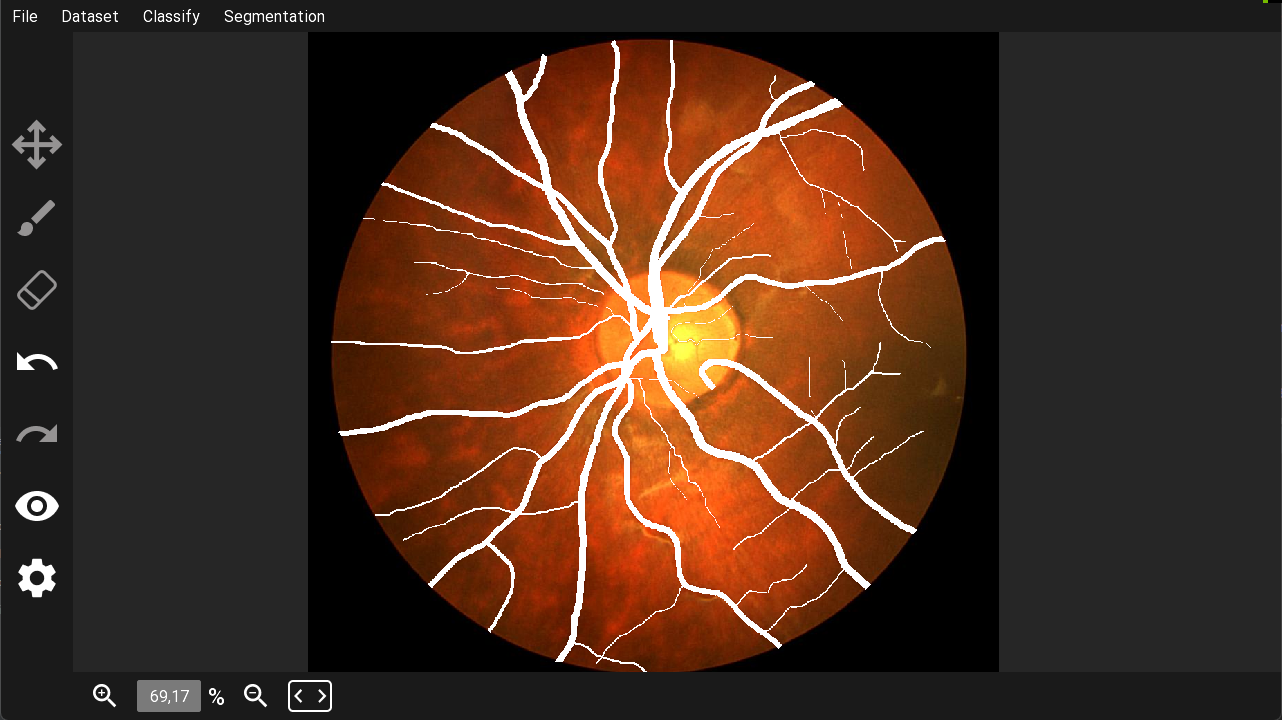}
\caption{Segmentação manual da Figura 1 na janela de edição de segmentações do software.}
\label{fig:soft1}
\end{figure}

Na prática, a vasculatura do olho é segmentada manualmente por profissionais da área, que é um processo mundano e laborioso. Além disso, requer perícia e um considerável grau de atenção e tempo \cite{Rodrigues2020}. Por esta razão, métodos automatizados de segmentação são importantes e podem ser obtidos por meio da aplicação de filtros de imagem e outras técnicas computacionais. 

\subsection{Segmentação por meio de processamento de imagem}
O software possui três algoritmos de segmentação implementados, o filtro de Frangi e dois filtros de segmentação baseados na técnica apresentada em \cite{Rodrigues2020}.

O filtro de Frangi \cite{Frangi} é bastante popular pela capacidade de aprimorar a visualização dos vasos sanguíneos em várias modalidades. Esta característica de maleabilidade para trabalhar com várias modalidades o faz uma ferramenta importante para a segmentação de imagens.

Inicialmente a imagem de entrada é convertida para escala de cinza utilizando a camada verde da imagem e depois o filtro de Frangi é aplicado.

Uma vantagem da aplicação do filtro de Frangi é a rapidez em gerar a segmentação. Entretando, a acurácia deste método é a mais baixa entre os algoritmos implementados.

Os filtros de conectividade \cite{Rodrigues2020} operam a partir do resultado obtido pelo filtro de Frangi. A ideia central é utilizar os pixels identificados pelo filtro de Frangi como pontos de semente, e a partir deles, aplicar a técnica de crescimento de região com a finalidade de explorar a característica de conectividade de um vaso sanguíneo. Assim, se um pixel pertence ao vaso, ele deve possuir pixels vizinhos que também pertencem à vasculatura do olho. A diferença entre os dois filtros é a região considerada para conectar os pixels. No primeiro filtro, é considerada uma região imediata ao redor do pixel semente, enquanto o segundo filtro considera uma região radial em torno do pixel.

\begin{figure}[h]
\centering
\includegraphics[width=\columnwidth]{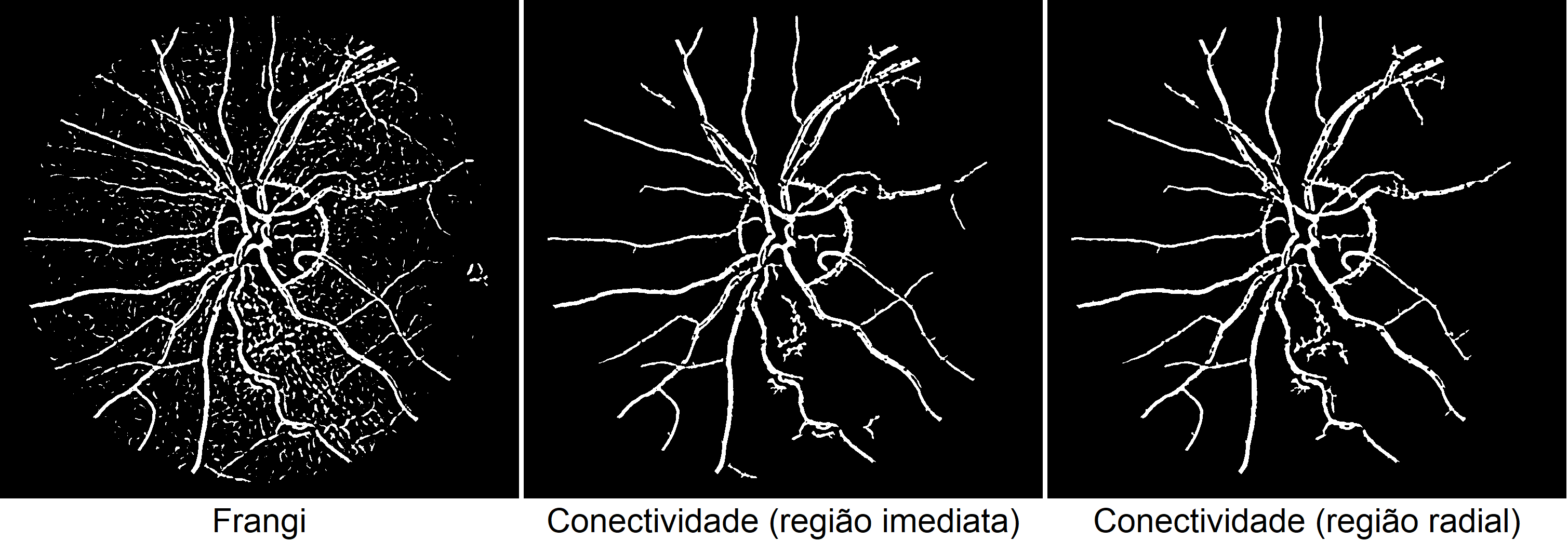}
\caption{Resultado da aplicação dos filtros a uma imagem de retina.}
\label{fig:results}
\end{figure}

%qual figura
Na Figura \ref{fig:results} temos o resultado da aplicação dos filtros do software para uma figura. Comparando os resultados obtidos, uma tendência que ocorre na segmentação utilizando o filtro de Frangi é a geração de ruídos na saída da aplicação. Esses ruídos podem ser removidos em parte, aplicando técnicas de redução de ruído como dilatação e erosão. Entretanto, algumas regiões da segmentação podem ser removidas também, gerando uma perda de informação. 

A aplicação dos filtros de conectividade geram resultados melhores, removendo os ruídos, ao considerar as regiões de forma conexa. Entretanto, o tempo de aplicação do filtro é maior. Seu resultado pode ser utilizado diretamente como a segmentação definitiva. Contudo, ainda podem ser necessários pequenos ajustes posteriores na segmentação.

Existem parâmetros configuráveis para cada um dos três filtros que podem ser alterados dentro do software para gerar um melhor resultado. 
\subsection{Classificadores e modelos preditivos}
Além da edição e dos filtros de segmentação, podemos criar conjuntos de dados e modelos preditivos de segmentação por meio do framework Weka \cite{Weka}.

A ideia central desta ferramenta é construir um modelo preditivo que receba características extraídas das imagens, construindo uma estrutura lógica de acordo com as entradas e saídas da classificação. A partir disto, temos um modelo que pode ser utilizado para criar uma segmentação a partir das características extraídas de novas imagens.

\begin{figure}[h]
\centering
\includegraphics[width=\columnwidth]{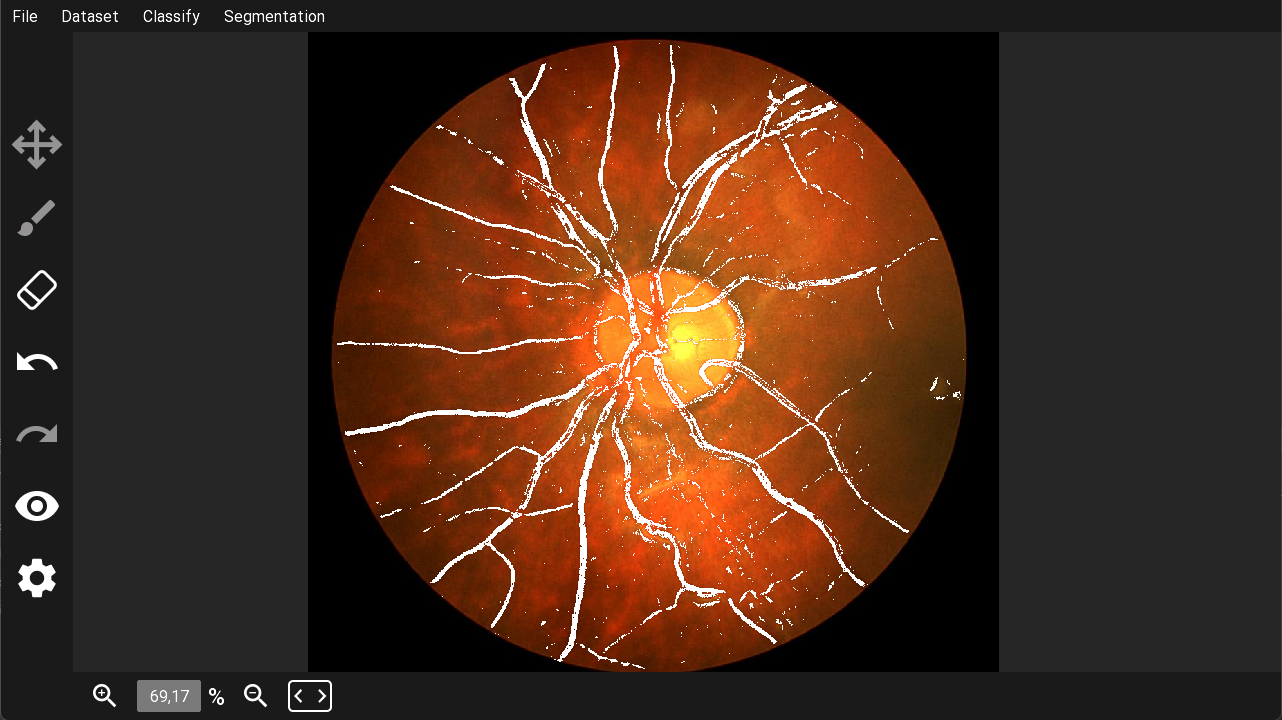}
\caption{Resultado da aplicação da classificação em uma imagem no software.}
\label{fig:class_result}
\end{figure}

Na Figura \ref{fig:class_result} temos o resultado de uma segmentação feita a partir dos classificadores do Weka. O resultado da classificação ainda pode ser futuramente melhorado. No momento, apenas três classificadores foram adicionados para testes das funcionalidades. Entretanto, como os classificadores já estão implementados na API Weka, basta adicioná-los nas opções do software.

O código fonte do software pode ser encontrado em \cite{github}. Algumas funcionalidades estão sendo implementadas e podem ser utilizadas parcialmente no momento. Também podem ser encontradas mais imagens do software, com foco nas ferramentas implementadas para edição das imagens, criação de datasets e modelos preditivos e aplicação dos filtros.

\section{Conclusões}

Este trabalho apresenta o desenvolvimento de um software multi-plataforma (mobile e desktop) para a criação e desenvolvimento de segmentações de imagens de retina, e automação dessas tarefas.

O software possui ferramentas para a criação de segmentações de forma manual e automatizada, e é o primeiro software de segmentações de retina que permite uma integração automática entre o uso de aprendizado de máquina e edição de imagens. 

O uso de filtros cria, de forma automatizada, imagens de segmentações para diversas modalidades de imagens de exames (fundo de retina, raios-x, etc), e com parâmetros ajustáveis, com a finalidade de melhorar os resultados obtidos.

A partir do aprendizado de máquina e com base na utilização do software, é possível automatizar ainda mais os processos, visto que, quanto mais dados são adicionados, os modelos podem ser subsequentemente retreinados gerando melhores acurácias na segmentação de novos vasos.

A versatilidade do software também é um ponto bastante importante de ser mencionado, pois permite que o usuário tenha várias opções para segmentar uma imagem, seja na criação de uma base de dados, para um exame de um paciente ou para o teste de uma técnica de segmentação.

Com isso, possíveis trabalhos futuros, incluem o aprimoramento das ferramentas de edição e adição de novos filtros de segmentação e suporte ao desenvolvimento à adição de conjuntos de features e filtros de maneira nativa no software.

\selectbiblanguage{brazil}
\bibliographystyle{ieeetr}
\bibliography{referencias}

\begin{thebibliography}{1}

\bibitem{morel2012variational}
J.-M. Morel and S.~Solimini, {\em Variational methods in image segmentation:
  with seven image processing experiments}, vol.~14.
\newblock Springer Science \& Business Media, 2012.

\bibitem{Rodrigues2016}
E.~O. Rodrigues, F.~F.~C. Morais, N.~A. O.~S. Morais, L.~S. Conci, L.~V. Neto,
  and A.~Conci, ``A novel approach for the automated segmentation and volume
  quantification of cardiac fats on computed tomography,'' {\em Comput. Methods
  Programs Biomedicine}, vol.~123, pp.~109–--128, 2016.

\bibitem{Rodrigues2018}
E.~Rodrigues, P.~Liatsis, L.~Satoru, and A.~Conci, ``Fractal triangular search:
  A metaheuristic for image content search,'' {\em IET Image Process}, vol.~12,
  no.~8, pp.~1475–--1484, 2018.

\bibitem{VargasCanas2012}
R.~Vargas-Canas and P.~Liatsis, ``Interactive retinal blood flow estimation
  from fluorescein angiograms,'' {\em JMech. Eng. Sci.}, vol.~226,
  pp.~2521–--2537, 2012.

\bibitem{retinopatia}
L.~Franco, ``Retinopatia diabética,'' {\em Acessado em: 28.06.2022.
  Disponível em: http://drleiserfranco.com.br/retinopatia-diabetica/}.

\bibitem{Rodrigues2020}
E.~O. Rodrigues, A.~Conci, and P.~Liatsis, ``Element: Multi-modal retinal
  vessel segmentation based on a coupled region growing and machine learning
  approach,'' {\em IEEE JOURNAL OF BIOMEDICAL AND HEALTH INFORMATICS}, vol.~24,
  pp.~3507--3519, 2020.

\bibitem{Weka}
E.~Frank, M.~A. Hall, and I.~H. Witten, ``The weka workbench, online appendix
  for “data mining: Practical machine learning tools and techniques”,''
  {\em Morgan Kaufmann, (Publishers, Inc.)}, 2016.

\bibitem{Frangi}
A.~F. Frangi, W.~J. Niessen, V.~K. L., and M.~A. Viergever, ``Multiscale vessel
  enhancement filtering,'' {\em Med. Image Comput. Comput.-Assisted
  Intervention}, vol.~1496, pp.~130--137, 1998.

\bibitem{github}
J.~H. Machado, ``Código fonte do software,'' {\em Acessado em: 28.06.2022.
  Disponível em: https://github.com/joaoHenriqueMachado/eyeSegment}.

\end{thebibliography}

\end{document}